\documentstyle[12pt]{article}
\begin{document}
\begin{center} 
{\Large\bf Direction Cryptography in Quantum Communications}
\vspace{.5in} 

{\bf Walter Simmons and Sandip Pakvasa\\} 
\vspace{.1in} 
 {\it 
Department of Physics and Astronomy, University of Hawaii at Manoa, 2505 Correa 
Road, Honolulu, HI 96822, USA \\}    

\date{\today}
\end{center}

\begin{abstract}

We examine a situation in which an information-carrying signal 
is sent from two sources to a common receiver.  The radiation travels 
through free space in the presence of noise.  The information resides 
in a relationship between the two beams.  We inquire into whether it 
is possible, in principle, that the locations of the transmitters can be 
concealed from a party who receives the radiation and decodes the 
information.  

Direction finding entails making a set of measurements on a
signal and constructing an analytic continuation of the time dependent
fields from the results.  The fact that this process is generally
different in quantum mechanics and in classical electrodynamics is the 
basis in this investigation.

We develop a model based upon encoding information into a 
microscopic, transverse, non-local quantum image (whose 
dimensions are of the order of a few wavelengths) and using a 
detector of a type recently proposed by Strekalov et al.  The optical 
system, which uses SPDC (Spontaneous Parametric Down 
Conversion), functions like a Heisenberg microscope: the transverse 
length, which encodes the signal information, is conjugate to the 
transverse momentum of the light.  In the model, reading the signal 
information spoils the directional resolution of the detector, while 
determining the directions to the sources spoils the information 
content.  Each beam, when examined in isolation, is random and 
indistinguishable from the background noise.

We conclude that quantum communications can, in principle, be made
secure against direction-finding, even from the party receiving the
communication.

\end{abstract}
\newpage
\section{Introduction}

 Classical cryptography had several weaknesses, including the need 
for either key distribution, or for the use of a public key system, whose 
algorithm could be subject to possible decipherment.  Other classical 
deceptive techniques, such as spread spectrum communications, 
padding the communications with dummy messages, etc., could help 
achieve better privacy, but it remained for quantum cryptography to 
remove the limitations on keys and make possible the encipherment 
of a signal that guaranteed privacy by the laws of quantum 
mechanics  \cite{4b, Bennett, 1b}.

In the case of signals sent through free space, it is often possible for a 
third party to direction-find the sources of quantum communications 
(or their concomitant classical signals), even if that eavesdropper cannot 
decipher the meaning of the signal.  Moreover, the signal's intended 
recipient, who has the necessary cryptographic key and equipment, 
generally cannot be deceived about the direction of propagation of the 
incoming radiation.

	Quantum mechanics is rich in physical phenomena, such as the 
superposition of amplitudes from various sources, EPR correlations, 
the interference of photons of different frequencies \cite{Larchuk},
etc., that might be used to 
engineer a system that could encrypt the direction of propagation of 
electromagnetic signals.  We undertake to investigate these 
possibilities in this work.

	When a single transmitter is involved, the direction to the 
transmitter can be determined by any party who can read information 
off of the transmitted signal, (even if the station is buried in noise), by 
simply measuring the recoil of the receiver.   This is true in 
both classical and quantum mechanics.

	We proceed to consider the case of two transmitters, 
in which the individual signals are pseudo-random, but which 
convey information to a distant party through a relationship 
between the two beams, which travel through a noisy environment.  
If the party receiving the signals cannot separate the signal 
from the noise, then the recoil of the apparatus points 
toward a position between the two transmitters and the 
individual sources cannot be separated.  Classically this 
will protect the sources from direction-finding from an 
eavesdropper only so long as he or she cannot decipher 
the signal.  Furthermore, it will not conceal the sources 
from the intended recipient who can identify and record 
the two components of the signal and determine the 
directions to the sources, in spite of the noise.

	Classically, ideal direction finding consists of measuring the 
electromagnetic fields throughout a volume of space for a period of 
time.  The measurements are made with arbitrarily high precision.  
Using analyticity, the full wave front is reconstructed and the positions 
of the sources are revealed.  With enough time and instrumentation, 
direction finding is always possible for classical fields, even in the 
presence of noise.  

	For the reasons just stated, we shall focus our analysis only upon 
quantum communications in which there are two sources and 
one receiver.  We shall assume the existence of an isotropic background noise.

\section{Direction Finding in Quantum Communications in Free Space}

	The introduction of quantum mechanics into communications 
widens the possibilities for cryptography very considerably.  
The properties of the electromagnetic fields can be divided, at least 
approximately, into two parts: the properties of the ensemble structure 
and the properties of the quanta.  The properties of the ensemble 
structure include the density of quanta in space and time (including 
the times and directions of transmission), and the inclusion of 
spurious quanta.  The quanta, for our purposes, are photons and biphotons.  

	As viewed from a quantum mechanical perspective, classical 
direction-finding consists of making ensemble measurements with sufficient 
statistical precision to enable the reconstruction of the space-time 
properties of the fields and subsequent location of the sources with 
desired accuracy. Methods such as spread spectrum communications 
vary a classical property (e.g. the frequency or time of transmission) 
but still send ensembles of identical quanta.  In other words, the designer 
breaks up large ensembles into smaller ones. 

	The essential departure that made quantum cryptography 
possible is based upon the observation \cite{4b} that individual quanta can not 
only carry information, but have many useful properties such as, fragility 
under measurement, superposition, and entanglement.  Of particular 
importance here, is the usual property that quantum states can be 
usefully measured only once.  

	In the model discussed here, all signal information is 
encoded into individual quanta (biphotons), which are transmitted 
as part of a heterogeneous ensemble.  The ensemble structure 
carries no signal information.  The biphotons carrying the signal
 information are transmitted at random intervals and noise 
quanta are added at will.  The intensity and other single 
particle statistics are relatively independent of the information
 content of the biphotons.  Thus, with wide latitude, the 
electromagnetic field can be constructed so that various 
distributions measured on the receiving end, whether by 
the intended receiver or by an eavesdropper, can generally be 
made to look like noise.  In a noisy background, the ensemble 
structure will carry no information useful for direction finding.

\section{Quantum Direction Finding}

	In 1989, Ou, Wang, and Mandel \cite{Ou} performed what may, in 
retrospect, have been the first experiment related to quantum 
ranging.  They showed that a positional ambiguity arises when two 
non-linear crystals are aligned along a common optical path.

	Giovannetti, Lloyd, and Maccone \cite{Giovannetti I} took up the 
question of cryptographic ranging in a one-dimensional situation and 
showed that a quantum mechanical system could ensure range 
privacy. 

	This work was stimulated by \cite{Giovannetti I}, but here, we are 
interested in direction cryptography in three-dimensional space.

\section{A Specific Problem in Direction Privacy}

	To make our analysis as simple as possible we pose a specific 
hypothetical problem for which we develop our model.  

	An agent living somewhere in the Galaxy desires to broadcast a 
signal to let some other intelligent beings elsewhere, know that she 
exists.  She is cautious, however, because she is concerned that a 
hostile being, which receives her signal, may direction-find her and 
make trouble.   Incidentally, such caution may offer a simple explanation
why no obvious signals from ETs (Extra-Terrestrials), have been yet
detected at the earth despite many searches; and thus provide a partial
explanation for the Fermi Question(or at least account for the great
silence \cite{Tarter}). Presumably ETs being smart as well as cautious, would
utilise the kind of direction cryptography discussed below, but most
likely much more advanced technological versions thereof.

	She notes that looking into even a small visual cone in the 
plane of the Galaxy reveals millions of stars, as well as weak 
background light and radio noise.  She desires to send a signal 
which, if studied with a direction finding apparatus, will look just like 
ordinary background noise, but which, if read properly, reveals some 
information.  

	She assumes that any agent worth contacting has already done 
his research, has set up all of his equipment, and will be ready to 
receive a signal at any time.  Moreover, she chooses to broadcast to 
limited groups of stars at a time.  

	We endeavor to engineer a communication system to solve this 
problem.

\section{Preliminary Discussion}

	Before taking up our model, in the next section, we discuss some aspects
of the problem semi-quantitatively here.

	Our problem is made easier by setting aside phenomena that 
do not occur in our hypothetical situation.  These include evanescent 
waves, back reaction on the transmitter, and the need for an 
informational cryptographic layer (we don't care who understands the 
message, we only desire to confuse those who do about its spatial origin.)

	The assumption that the receiving agent is well prepared, 
allows us to radiate only a limited amount of total energy and a limited 
number of quantum states, all of which might be different from one 
another; therefore, useful high statistics measurements cannot be 
made by the receiver.  The assumption that the radiation is targeted 
at a limited volume allows us to choose radiation with a small signal 
strength (or a limited coherence length) away from the optical path, 
so that wide-angle measurements, made with detectors separated by 
astronomical distances, will yield no information.  Moreover, because 
of these assumptions, the range to the receiver will be approximately 
known, allowing convenient conversion of approximate distances into 
angular measures.

	We desire that the signal be readable with an instrument, which 
measures photon pairs originating from different directions in space, 
but not with an instrument that detects the photons from the individual 
sources.  This corresponds to a choice of large transverse 
momentum acceptance, versus small transverse momentum acceptance.  
Utilizing the Uncertainty Principle, we choose to engineer information 
into the signals in the form of an image, of microscopic transverse 
dimensions.  The radiation will be transmitted from a pair of radiators 
separated by an astronomical distance.

	As we discuss in a section, below, the transference of images 
into quantum correlations between optical beams has been studied 
for several years.  Recently, using Spontaneous Parametric Down 
Conversion, clear images of letters, arrayed transversely to the beam 
directions, have been encoded in correlations and subsequently 
recovered from individual photon detection measurements
\cite{Group B}.

	For our purposes, we require an image which is microscopic in 
size.  The minimum dimension is of the order of one wavelength, as 
smaller images do not propagate.  To date, there have been no 
experimental demonstrations of the encoding and decoding of such 
small images, while there is no problem of principle in doing such 
experiments. See also the Appendix.  

	The detection of quantum correlations over ranges of tens of 
kilometers has been demonstrated \cite{Group C}.  We shall assume that 
the detection can be extended to astronomical ranges.

	In our hypothetical apparatus, we encode a small transverse 
image onto a pair of optical beams, prepared by Spontaneous 
Parametric Down Conversion \cite{Group B}.  The beam is split and directed 
toward reflectors located far apart (the distance L, which is of the order
of several parsecs), where the two beams 
are directed at the receiving party, located at a known distance, R.  
The form of the image is unimportant and may be as simple as some 
dots separated by a distance, $Y_0$.

	Let us call the opening angle, (at the receiver), between the two 
transmitters, $\Delta \Theta_0$, and call a typical smaller opening angle, such as 
might be necessary to isolate the radiation from one transmitter from 
the radiation from the other transmitter, $\Delta \Theta$.  With each of these 
angles, we associate a spread of transverse momentum.
\begin{equation}
\Delta P_{y,0} \sim P_Z \Delta \Theta_0 \ \ \mbox{and} \ \ 
\Delta P_{y} \sim P_Z \Delta \Theta
\end{equation}	    

Note that the momentum along the direction of propagation, $P_Z$ , 
can be measured independently of the angles.  Thus a measurement 
of one of the angles (equivalently, restricting the radiation to within 
that angle) amounts to a measurement of the transverse momentum.

	We want to engineer the image so that it can be viewed with 
quantum mechanical measurements only when no precise 
measurement or strong restriction is made upon the transverse 
momentum of either photon.  When the radiation from either 
transmitter is isolated, we want the Uncertainty Principle to prevent 
the observation (and recording) of the image information in the 
radiation.  We choose $Y_0$ such that,
\begin{equation}
Y_0 \gg 
\frac{1}{\Delta P_{y,0}} =
\frac{\lambda}{4 \pi \Delta \Theta_0} \ \
	 	\mbox{but also,} \ \
Y_0 \ll
\frac{1}{\Delta P_{y}} =
\frac{\lambda}{4 \pi \Delta \Theta}
\end{equation}
	  
In order that the radiation propagate, we must also have,
\begin{equation}
Y_0 \geq  O (\lambda)
\end{equation}			 

	These conditions are not difficult to achieve if L and R are of 
astronomical sizes.  Therefore, we can engineer the transmitter and 
image correlations so that the receiver must use a large acceptance 
angle instrument in order to be able to extract the image.  In that 
case, the recoil momentum measured points back toward a position 
between the transmitters and gives very little information about their 
true location.  If the receiver measures the transverse momenta of the 
individual photons with sufficient precision to determine the position 
of their respective transmitters, then the Uncertainty Principle implies 
that the information will be washed out and the photons are 
indistinguishable from background noise.

\section{Non-Local Quantum Image Based Model}

It has been known for some time that the quantum correlations between 
particle states, originally discussed by EPR, can carry information.  In 
Spontaneous Parametric Down Conversion (SPDC), a biphoton state 
is created by the interaction of a laser beam with a non-linear crystal \cite{MW}.  
Within a first order perturbation theory treatment, (with the pump beam
treated classically), the interaction 
Hamiltonian has a factor.
\begin{equation}
H_I \rightarrow a^+_1 a^+_2 a_p
\end{equation}			 
where (1) and (2) refer to outgoing photon states and (p) refers to 
the laser beam, which is generally treated classically.  From this 
we see that,
\begin{equation}
exp (-iH \ _It) |  0 >_1 |0>_2 \ \rightarrow \sum_{n}\ C_n (t) | n >_1 | n >_2
\end{equation}
and hence, the state consists of two highly entangled multi-photon states 
with perfectly correlated photon number.  The states of greatest interest to 
date are the biphoton states, consisting of an entangled pair of photons.  
Within a first order perturbation context these can be written as single photons 
states which are entangled in momentum. The two modes are 
traditionally called signal (S) and idler (I).  
Taking these to be continuous, the Hamiltonian, above, 
evolves the vacuum into the state,
\begin{equation}
|  Biphoton > \ \rightarrow  \int dk_1 \int dk_2 \Phi (k_1, k_2, k_p) \mid k_1, 
\sigma_1 >_I | k_2, \sigma_2 >_S
\label{vectors}
\end{equation}

A biphoton is an excitation of the electromagnetic field, which need 
not be represented perfectly in perturbation theory.  For example, the biphoton 
can carry the same k vector as the pump beam.  This was noted theoretically 
by \cite{Lukin, Physics} and has recently been observed experimentally by \cite{Physics}.  
Strekalov et al. \cite{Strekalov} have discussed the point that the
biphoton has properties 
which are not described by perturbation theory.

Just as in EPR states, information encoded in, say, momentum or 
polarization relationships between the S and I photons, resides in a 
relationship between those properties of the individual photons.  For 
this reason, this kind of information is often called quantum 
non-local information.

A special case of Eq.~\ref{vectors}, in which the 
function $\Phi$ depends only upon the difference 
of the transverse momenta, could be 
\begin{equation}
|  Biphoton > \ \rightarrow  \int dq_1 \int dq_2 \Phi (q_1 - q_2)| q_1>_I |q_2>_S
\label{vector}
\end{equation}
where the $q$ vectors are the transverse momenta of the photons and where other 
variables have been suppressed.  

If we engineer the function  to be peaked around zero, then we observe the 
following properties of the state in Eq.~\ref{vector}:

a.)  A measurement of either single photon state, (S or I), results in a 
random result for the corresponding q vector.

b.)  Following the first measurement, the possible momentum of the other single 
photon state, (I or S), is sharply peaked around the same value of q.

We know from classical optics \cite{Goodman} that when light is passed through an aperture of 
diameter w and at a range z from the source, (w/z is an angular measure), 
then the spatial frequency is cut off at 
\begin{equation}
\frac{w}{z \lambda}
\label{cutoff}
\end{equation}
			 							
It is usually assumed that classical transfer functions for passive 
optical devices carry over to the quantum mechanical description.  
Thus, if one of the photons in Eq.~\ref{vectors}, (S or I), is passed through 
an aperture of sufficiently narrow width, its transverse momentum 
will be constricted to a value that can pass through the aperture.  
It follows from the properties listed above, that the amplitude 
for the other photon (I or S) is similarly reduced.  

Our model for communication with direction privacy consists 
of encoding information on large values of 
the transverse momentum 
variables  $(q_1$ or $q_2)$ of biphotons, which consist, 
generally of photons of unequal
frequencies.  As per property (a), a measurement of an 
isolated photon yields a noise value.  An attempt to restrict the transverse 
momentum of either photon in order to narrow the direction of 
propagation into an angle determined by w/z, results in the destruction 
of the transverse momentum information in both photons.

A measurement of the biphoton as a single quantum state causes 
the measuring apparatus to recoil in the direction of motion of the 
biphoton and the individual photon directions of motion are lost.

Obviously, it is essential that it be possible to extract the transverse 
information from the biphoton.  For that reason, we shall briefly 
discuss the status of transverse, quantum non-local images.

Within the past few years, it has become possible to 
convey images using biphoton states.   Investigations of fringe 
visibility and the transverse spatial structure of SPDC light were 
carried out starting around 1993 \cite{Group B, Ribeiro}. Two early experiments 
are those of Strekalov et al. 1995 \cite{Group B} and Pittman et al. 
1995 \cite{Group B}.  
Some theoretical discussion is given in \cite{Theory A}.  Fonseca et
al. \cite{Theory A} 
demonstrated that the angular spectrum of the pump beam can 
be recovered from the fourth order correlations of the signal 
and idler fields.  It has recently been demonstrated theoretically 
and experimentally \cite{Nogueira} that these images have photon anti-bunching 
in spatial variables transverse to the direction of biphoton propagation, 
thus extending the concept of temporal anti-bunching and demonstrating 
that the correlations in the transverse plane have no classical analog. 

Our purpose in choosing SPDC is to have a model process.  
The state of technology is not yet such that we can use SPDC in 
its present state for direction cryptography.  The realm of microscopic
 images has not been studied experimentally, and our model requires 
a new kind of biphoton detector such as that proposed by Strekalov 
et al.  \cite{Strekalov}.  Lukin et al. \cite{Lukin} have discussed the theory of the 
detection of entangled photons in connection with the phenomenon 
of storage of light in atomic vapor. 

\section{Other Direction Finding Approaches}

In this section we consider a specific attack on the cryptography 
and add some comments about the ensemble properties.

Let us suppose that the receiver attacks the privacy as follows.  
He focuses an image of the sky onto a plane in his laboratory.  
Behind this plane, he places a biphoton detector.   He then blocks 
out a small section of the image at the (guessed) position of the image 
of one of the two sources.  When the signal in the detector vanishes, he has 
obviously covered up the image of the source.  

From Eq.~\ref{cutoff}, we see that the image of each source must be broad 
enough to encompass at least the width of the aperture we considered 
earlier.  The angular width of the source images are greater than the 
angular separation of the two sources.  Thus, this method does not work.	

In an attack, in general, the receiver places biphoton detectors 
throughout a volume of space and records, for each biphoton event at 
each detector, the arrival time, the total momentum, polarization, and, 
of course, the transverse displacement data (the microscopic image), which 
carries the signal information.  He next attempts to use this data to 
derive the source directions.  

We emphasize that the signal information is a property of 
individual biphotons, and the ensemble characteristics can be 
engineered for maximum deception at the transmitter.  We also 
emphasize that each biphoton can be usefully measured only once, 
never at several sets of detectors.  Correlations between count rates by 
different detectors generally measure ensemble characteristics.  

The introduction of an image distribution into an ensemble of 
biphotons like those in Eq.~\ref{vectors} can be expected to bias the 
single photon statistics.  However, the bias can be calculated and 
offset by introducing single photon noise. 

If the microscopic transverse image is static, then the biphotons 
can arrive in any sequence.  The transmitter emits them a few at a 
time and at random intervals.  We have already discussed the idea 
that the structure of the quantum signal in the transverse direction 
can be chosen to be random on macroscopic scales.  Therefore, 
neither the time of arrival data, nor the macroscopic spatial 
distribution data, contains significant direction information.

\section{Conclusions}

	We have investigated whether quantum communications can be 
made resistant to direction-finding.

	We considered a model in which a heterogeneous ensemble of 
biphotons carries information, in the form of a microscopic transverse image, 
to an intended receiver.  We conclude that it is possible to engineer the 
radiation field so that even the receiver, who knows how to read the
informational image, cannot determine the precise angular position of
sources from which the biphotons were transmitted.

In this paper, we have raised the possibility that quantum 
communications can be made secure against direction finding; 
and have suggested specific techniques to achieve this goal.

Many reasons have been proposed to explain the fact that SETI 
researchers have detected no signals from ETs (the so-called Fermi
question).  For a recent review, 
see \cite{universe}. The model described here provides the germ of a solution 
to this problem.  When technology has developed further than it stands 
today, it may be possible to design a plausible quantum search strategy 
for use in SETI.

\section{Acknowledgements}

	We thank Xerxes Tata for many clarifying discussions.  This 
work was partially supported by the U.S. Department of Energy under 
grant number DE-FG-03-94ER40833. 


\section{Appendix:  Information Propagation of Biphotons}	

As is well known \cite{Goodman}, an electromagnetic wave cannot 
propagate if the transverse wave vector, q, exceeds a bound 
determined by the relationship,
\begin{equation}
\hat{q}^2 + k^2_z = \left ( \frac{2 \pi}{\lambda} \right )^2
\end{equation}
		 
The pump beam has half the wavelength of the individual 
outgoing photons.  Therefore, the maximum range of wave vector for 
each daughter photon is, in perturbation theory, lower than that of the 
pump photon.  The transverse momentum in the disallowed range 
can, however, be carried as a net momentum of the pair of photons, 
depending upon the absorption of momentum in the crystal.  
Moreover, if the biphoton is regarded as a single quantum state, it 
has, in principle, the capability of carrying the same range of 
transverse momenta as the pump state from which it arises.  This 
behavior has been discussed by Fonseca et al. 
\cite{Fonseca}, and by Jacobson, et. al. \cite{Jacobson}.  This has
been studied experimentally by Edamatsu, et al. \cite{Physics}.  See also \cite{Rubin}.

The interesting point here is that when the photons are treated 
individually, they cannot carry all of the transverse momentum 
information of the pump photons.  However, the pair can carry all of 
the information in the pump beam.

\end{document}